\documentclass[twocolumn,showpacs,floats,floatfix,superscriptaddress,aps,pra]{revtex4-1}
\usepackage{amsfonts}
\usepackage{amssymb}
\usepackage{amsmath}
\usepackage{calc}
\usepackage{graphicx}
\usepackage{bm}

\usepackage[normalem]{ulem}
\usepackage{amsmath,amssymb}
\usepackage{multirow}
\usepackage{xcolor,soul}
\usepackage{srcltx}
\usepackage{hyperref,graphicx}

\def\be{ \begin{equation}}
\def\ee{ \end{equation}}
\def\bea{ \begin{eqnarray}}
\def\eea{ \end{eqnarray}}
\def\bse{ \begin{subequations}}
\def\ese{ \end{subequations}}
\def\bc{ \begin{center}}
\def\ec{ \end{center}}

\begin{document}

\author{Stefano Longhi$^{*}$} 
\affiliation{Dipartimento di Fisica, Politecnico di Milano and Istituto di Fotonica e Nanotecnologie del Consiglio Nazionale delle Ricerche, Piazza L. da Vinci 32, I-20133 Milano, Italy}
\email{stefano.longhi@polimi.it}

\title{Probing topological phases in waveguide superlattices}
  \normalsize

%\date{.}

%
\bigskip
\begin{abstract}
\noindent  
One-dimensional superlattices with modulated coupling constants show rich topological properties and tunable edge states. Beyond the dimeric case, probing the topological properties of superlattices is a challenge. Here we suggest a rather general method of bulk probing topological invariants in waveguide superlattices based on spatial displacement of discretized beams. A judiciously tailored initial beam excitation of the lattice, corresponding to superposition of Wannier functions,  provides a direct measure of the band gap topological numbers. For  a quadrimeric superlattice, a simple bulk probing method, which avoids Wannier states, is proposed to discriminate the existence of zero-energy topological edge states from non-topological ones.
\end{abstract}

%\pacs{03.65.-w,  72.20.Ee, 72.15.Rn,  73.43.-f, 71.30.+h }

% 03.65.-w Quantum mechanics
% 72.20.Ee    Mobility edges; hopping 
%42.25.Bs, 42.82.Et, 11.30.Er 
% 11.30.Er 	Charge conjugation, parity, time reversal, and other discrete symmetries
% Supersymmetry,
% 72.15.Rn: Localization effects
%71.30.+h Metal-insulator transitions and other electronic transitions
%  73.43.-f Quantum Hall effects

\maketitle
 
 {\it Introduction.} In recent years, topological photonics has emerged as a new
paradigm to robustly manipulate light propagation under continuous
deformations and to emulate in optics topological phases of matter (see \cite{r1,r2} and references therein).
Topological phases are described by global invariants, such as the Zak phase and Chern numbers.
Since topological invariants describe non-local properties of the system, ways to relate topological invariants with measurable quantities is of major relevance both theoretically and experimentally. The bulk-boundary correspondence \cite{r3,r4}, relating the bulk topological invariants with the number of edge states in finite systems with open boundaries, 
provides the simplest route to measure topological numbers \cite{r4bis,r5,r6,r7,r7bis,r8}. However, edge states without topological protection could coexist with topological ones, making edge measurements ambiguous in certain cases. Bulk probing of topological invariants has been proposed and demonstrated in a series of experiments with synthetic matter, such as those based on Bloch oscillations \cite{r9,r10}, unitary or non-unitary quantum walks (QWs) \cite{r11,r12,r14,r15,r16,r17,r18,r19}, and topological pumping \cite{r20,r21}.
 In one-dimensional (1D) systems the bulk topological properties of the Bloch bands
are characterized by the quantized Zak phase \cite{r22,r23,r24,r25,r26}. In two-band systems, such as in celebrated Su-Schrieffer-Heeger (SSH) model \cite{r3}, beam displacement in a photonic QW provides a simple and powerful approach to detect the Zak phase in the bulk \cite{r15,r16,r17,r18,r19}. Multi band superlattices show richer topological features than two-band models, such as the ability to tune the number of topological edge states by controlling the inter-dimer coupling \cite{r26}. In such systems, topological edge states can coexist with non-topological ones \cite{r25}, making topological phase characterization of the edge ambiguous. Recently, the displacement method \cite{r17} has been extended to measure a global winding number in chiral superlattices \cite{r27bis}, however an open question is whether photonic QWs can provide a measure of topological numbers {\em in each gap} of the superlattice or can be used when the system does not show chiral symmetry.\\
In this Letter it is shown that photonic QWs of judiciously-tailored discretized optical beams in waveguide lattices, corresponding to a superposition of Wannier functions, provide a general route to measure the Zak phase of various superlattice bands. In case of a quadrimeric superlattice, a simple setup is suggested  to detect particle-hole zero-energy modes \cite{r27} and to discriminate them from non-topological edge states. \par
{\it Topological properties of superlattices and edge states.}  
We consider a 1D waveguide superlattice with open boundary conditions comprising $N$ unit cells, each unit cell hosting $M$ equal waveguides, with inter-cell coupling constant $\tau$ and intra-cell couplings $t_1, t_2,..., t_{M-1}$ (Fig.1). The different coupling constants can be readily controlled by adjusting waveguide spacing (see, for instance, \cite{r28,r29}). The Hamiltonian of the system reads
\begin{equation}
\hat{H}=\sum_{n=1}^N \left\{ \left( \sum_{l=1}^{M-1} t_l \hat{a}^{\dag}_{n,l} \hat{a}_{n,l+1}\right) + \tau \hat{a}^{\dag}_{n,M} \hat{a}_{n+1,1} + H.c. \right\}
\end{equation}
where $\hat{a}^{\dag}_{n,l}$ is the photon creation operator at the $l-th$ waveguide of the lattice in the $n-th$ unit cell. For $M=2$, this model reproduces the celebrated two-band SSH model.
In bulk momentum space representation, the Hamiltonian is described by the $M \times M$ matrix $\mathcal{H}(k)$ defined by
\begin{eqnarray}
\mathcal{H}_{n,m}(k) & = &  t_n \delta_{n,m-1}+t_m \delta_{n,m+1}+ \tau \exp(-ik) \delta_{n,1} \delta_{m,M} \nonumber \\
& + & \tau \exp(ik) \delta_{n,M} \delta_{m,1}
\end{eqnarray}
where $k$ is the Bloch wave number, which varies in the interval $(-\pi,\pi)$. The $M$ energy bands $E_l(k)$ and corresponding Bloch functions $u_l(k)$ of the superlattice $(l=1,2,..,M$) are defined by the eigenvalue equation $\mathcal{H}(k) u_l(k)=E_l(k) u_l(k)$, with the normalization condition $ \langle u_l(k) | u_n (k) \rangle= \delta_{n,l}$. Owing to the particle-hole symmetry \cite{r26,r27}, the energy spectrum appears to be mirror symmetric with respect to the zero energy $E=0$. The topological properties of the $(M-1)$ band gaps are characterized
by the sum of Zak phases of all the isolated bands below the corresponding band gap \cite{r22,r23,r25,r26}, i.e. by the numbers
\begin{equation}
\mathcal{N}_n= \gamma_1+\gamma_2+...+\gamma_n 
\end{equation}
where $n=1,2,..,M-1$ is the band gap number (ordered for increasing values of energy) and $\gamma_l \equiv i \int_{-\pi}^{\pi} dk \langle u_l(k) | (du_l / dk) \rangle$ is the Zak phase of the $l-th$ band. As shown in previous works \cite{r22,r23,r26}, the Zak phase and thus the gap numbers $\mathcal{N}_l$  are defined apart from integer multiplies than $ 2 \pi$, depend on the choice of the origin in the unit cell, and they are not quantized rather generally. However,  when the system possesses inversion symmetry, i.e. when $t_{M-l}=t_l$, the Zak phase is quantized and can take two possible values: $0$ or $\pi$ \cite{r23,r26}. Correspondingly, 
the gap numbers $\mathcal{N}_l$ are topological and quantized, taking the two possible values $0$ or $\pi$ (mod $ 2\pi$). Interestingly, in the gaps with topological number $\pi$, nearly-degenerate topological edge states, localized one on the left (L) and the other on the right (R) edges of the lattice, do exist \cite{r26}.\\ 
The eigenenergies and domain of existence of edge states can be determined, in the most general case, by extending the procedure earlier introduced in \cite{r29} for surface bound states in the continuum. Indicating by $\mathcal{H}_L$ the $(M-1) \times (M-1)$ matrix defined by $(\mathcal{H}_L)_{n,m}=t_n \delta_{n,m-1}+t_m \delta_{n,m+1}$ and by $E_{L,\sigma}$, $v^{(L, \sigma)}$ the eigenvalues and corresponding eigenvectors of $\mathcal{H}_L$, i.e. $\mathcal{H}_L v^{(L, \sigma)}=E_{L, \sigma} v^{(L, \sigma)}$, it can be shown that there are up to $(M-1)$ left edge states with energies $E_{L, \sigma}$; moreover, the left edge state with energy $E_{L, \sigma}$ does exist provided that $\tau>t_{M-1} |v_{M-1}^{(L, \sigma)}/ v_{1}^{(L, \sigma)} |$. Likewise,  let us indicate by $\mathcal{H}_R$ the $(M-1) \times (M-1)$ matrix defined by $(\mathcal{H}_R)_{n,m}=t_{M-n} \delta_{n,m-1}+t_{M-m} \delta_{n,m+1}$, and by $E_{R,\sigma}$, $v^{(R, \sigma)}$ the eigenvalues and corresponding eigenvectors of $\mathcal{H}_R$, i.e. $\mathcal{H}_R v^{(R, \sigma)}=E_{R, \sigma} v^{(R, \sigma)}$. Then, it can be shown that there are up to $(M-1)$ right edge states with energies $E_{R, \sigma}$; moreover, the right edge state with energy $E_{R, \sigma}$ does exist provided that $\tau>t_{1} |v_{M-1}^{(R, \sigma)}/ v_{1}^{(R, \sigma)} |$. Note that, in the presence of inversion symmetry, one has $\mathcal{H}_{L}=\mathcal{H}_{R}$ and thus the energies of left and right edge states become degenerate: this mean that in superlattices with inversion symmetry the edge states appear in pairs at both left and right edges with the same energy.\\
Figure 2 shows an example of numerically-computed energy spectrum (left panels) and corresponding gap numbers $\mathcal{N}_n$ (right panels) versus the inter-cell coupling $\tau$ in two superlattices comprising $M=4$ waveguides in each unit cell. In Fig.2(a) there is inversion symmetry ($t_3=t_1$), and the gap numbers $\mathcal{N}_n$ are quantized for all gaps. Note that degenerate left and right edge states appear in the middle gap (energy $E=0$) and in the upper/lower gaps (energies $\pm \sqrt{t_1^2+t_2^2}$) as $\tau$ is increased above the two values $\tau_1=t_1^2/t_2$ and $\tau_2=t_2$, respectively. The topological nature of the edge states, signaled by the quantization of the gap numbers, can be also revealed by the fact that, as $\tau$ is varied, the edge  states appear and disappear as the band gaps close and reopen. Figure 2(b) corresponds to a superlattice without inversion symmetry. In this case the gap numbers in the upper and lower band gaps are not quantized and vary continuously  as the inter-cell coupling $\tau$ is varied. Correspondingly, L and R edge states observed by increasing $\tau$ are not topological (they emanate from the bands without gap closing) and are not degenerate in energy. Interestingly, for the middle gap the number $\mathcal{N}_2$ is quantized and topological zero-energy  degenerate L and R edge states are observed by increasing $\tau$ after band gap closing and reopening at $\tau=\tau_0 \equiv t_1  t_3 / t_2$ [Fig.2(b)]. The born of such edge states corresponds to a sudden jump of the gap number $\mathcal{N}_2$ from trivial (0) to nontrivial ($\pi$) values.  This means that in a superlattice with broken inversion symmetry topological and non-topological edge states can coexist.\par

  \begin{figure}[htb]
\centerline{\includegraphics[width=8.4cm]{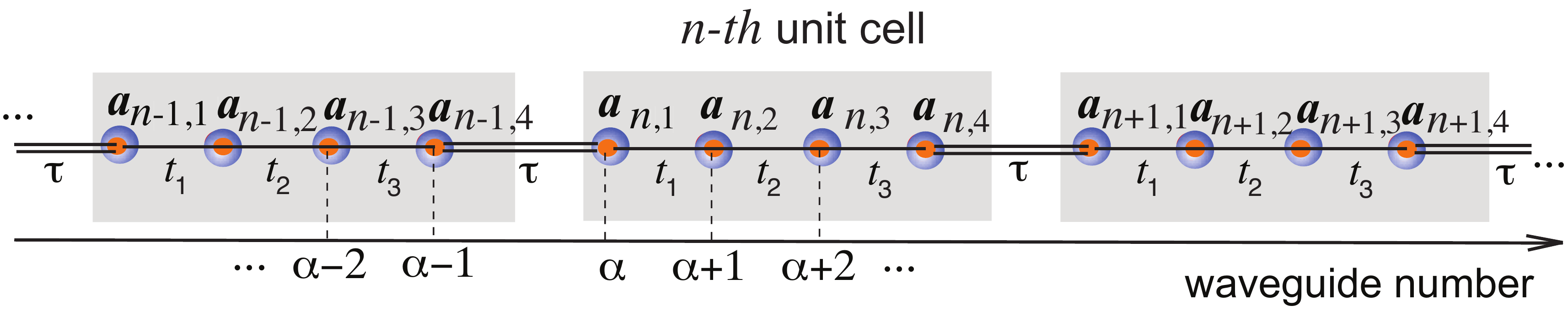}} \caption{ \small
(Color online) Schematic of a waveguide superlattice. Each unit cell contains $M$ equal waveguides ($M=4$ in the figure). Inter-cell coupling is $\tau$, whereas $t_1$, $t_2$, ..., $t_{M-1}$ are intra-cell couplings. For an inversion-symmetric superlattice, the condition $t_l=t_{M-l}$ holds.}
\end{figure} 
\begin{figure}[htb]
\centerline{\includegraphics[width=8.5cm]{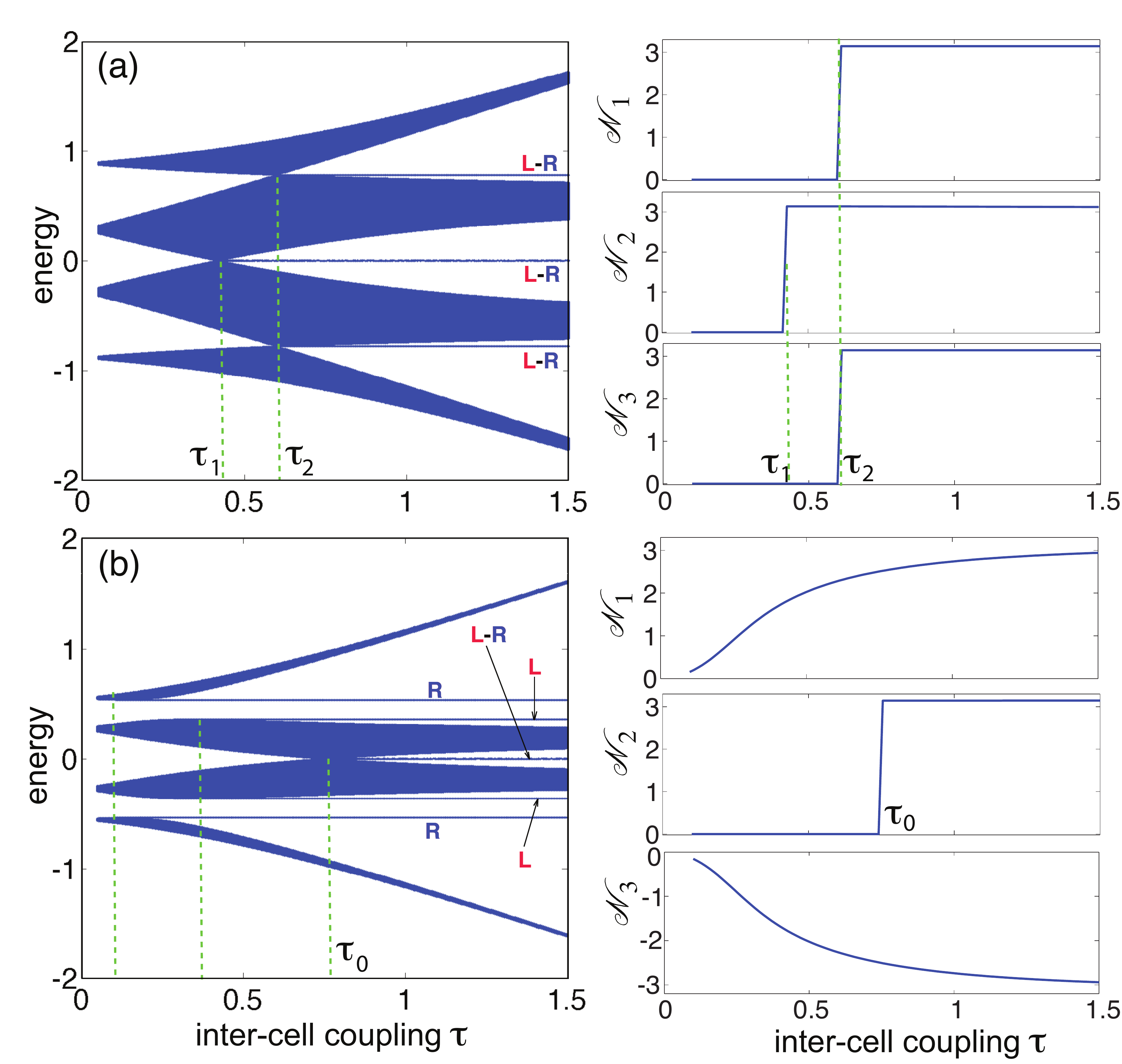}} \caption{ \small
(Color online) Energy spectrum (left panels) and corresponding behavior of the sum of Zak phases $\mathcal{N}_n=\sum_{l=1}^{n} \gamma_l$ (right panels) versus inter-cell coupling $\tau$ in a quadrimeric superlattice ($M=4$) (a) with inversion symmetry ($t_1=t_3=0.5$, $t_2=0.6$.) and (b) without inversion symmetry ($t_1=0.3$, $t_2=0.2$, $t_3=0.5$). A lattice with open boundary conditions comprising $N=200$ unit cells has been assumed in the calculation of the energy spectrum. Edge states, localized either at the left (L) or right (R) edges of the lattice, are clearly visible in both (a) and (b). The dashed vertical lines in the left panels highlight the threshold values $\tau$ above which the edge states appear. In (a) all edge states are topological, whereas in (b) only the zero-energy state in the central gap is topological.}
\end{figure}
{\it Wannier function excitation, beam displacement and gap numbers.}  Recent works have shown that, in a two-band lattice such as in the SSH model, the average beam displacement in a photonic QW with single-site  initial excitation provides a direct measure in the bulk of Zak phase \cite{r15,r17,r18,r19}. A main open question is whether  photonic QSs in a multi-gap superlattice can be related to the Zak phases and gap numbers of the various bands/gaps. The propagation of the field amplitudes in the various waveguides can be written as a superposition of Bloch modes
\begin{equation}
a_n(z)= \sum_{l=1}^{M} \int_{-\pi}^{\pi}dk C_l(k) u_l(k) \exp [ikn-iE_l(k)z] 
\end{equation}
where we have set $a_n \equiv (a_{n,1}, a_{n,2},...,a_{n,M})^T$ and where the spectral amplitudes $C_l(k)$ are determined by the initial excitation condition $a_n(0)$ of the array according to 
\begin{equation}
C_l(k)=(1/ 2 \pi) \sum_n \langle u_l(k) | a_n(0) \rangle \exp(-ikn)
\end{equation}
with the normalization condition $ 2 \pi \sum_l \int_{-\pi}^{\pi}dk |C_l(k)|^2=1$. We are interested to calculate the integral mean displacement of the discretized beam along the propagation distance $z$, defined by
\begin{equation}
\mathcal{D}(z)= (1/z) \int_0^z d \xi \sum_{n=-\infty}^{\infty} n \langle a_n( \xi) | a_n(\xi) \rangle
\end{equation}
which is a measurable quantity. Substitution of Eq.(4) into Eq.(6), after some calculations it can be shown that, in the large $z$ limit, $\mathcal{D}(z)$ reaches the asymptotic value
\begin{equation}
\mathcal{D}_{as}=2 \pi i \sum_{l=1}^M \int_{-\pi}^{\pi} dk \left( C_l^*(k) \frac{dC_l}{dk}+|C_l(k)|^2 \langle u_l(k) | \frac{du_l}{dk} \rangle \right).
\end{equation} 
Let us assume that $C_l(k)$ is independent of $k$, i.e. let us assume uniform band filling. For $C_l(k)=(1/ 2 \pi) \delta_{l,n}$, i.e. when the $n-th$ Bloch band solely is uniformly excited at input plane, the initial beam excitation $a_n(0)=(1 / 2 \pi) \int_{-\pi}^{\pi} dk u_n(k) \exp(ikn)$ reproduces the Wannier function of the $n-th$ band \cite{r30} and correspondingly one has $\mathcal{D}_{as}=\gamma_n / (2 \pi)$: this is a well-known result, that relates Zak phase to the Wannier function displacement in a crystal \cite{r30}. The most interesting case is attained when we excite uniformly the lowest $S$-bands of the superlattice. By assuming $C_l(k)=1/ (2 \sqrt{S} \pi)$, i.e. $a_n(0)=(1 / 2 \sqrt{S} \pi) \sum_{l=1}^{S} \int_{-\pi}^{\pi} dk u_l(k) \exp(ikn) \equiv A_n^{(S)}$, from Eq.(7) it follows that
\begin{equation}
\mathcal{D}_{as}=(1/2 \pi S) \sum_{l=1}^S \gamma_l= \mathcal{N}_S /(2 \pi S)
\end{equation}
i.e. a measure of the mean beam displacement yields the topological number $\mathcal{N}_S$ of the $S-th$ band gap, scaled by the factor $ 2 \pi S$. Figure 3 shows examples of numerically-computed evolution of integral beam displacement in a quadrimeric superlattice with [Fig.3(a,b)] and without [Fig.3(c)] inversion symmetry, corroborating the theoretical prediction of Eq.(8). In the figure, the amplitudes of the excitation field at the input plane $z=0$ is shown in the insets, while propagation distancee $z$ is normalized to the inverse of the inter-cell coupling constant $\tau$. In all cases the initial field distribution  turns out to be real and requires to excite some waveguides ($ \sim 10-30$) with tailored intensity and phase ($0$ or $\pi$ depending on the sign of $a_n$). Intensity control can be provided by proper beam shaping methods (for example using a spatial light modulator \cite{r31} or a binary phase plate \cite{r32} and a Fourier filter), while $\pi$ phase shifts could be realized by waveguide segmentation \cite{r33}. Note that the asymptotic value $\mathcal{D}_{as}$ is reached after a propagation length $ z \sim 40/ \tau$. For example, for case of Fig.3(a) and for a typical value $\tau \sim 3 \; {\rm cm}^{-1}$ in waveguide lattices manufactured by fs-laser writing \cite{r29,r31},  such a distance corresponds to a sample length $L\sim 13$ cm, which is feasible with current technology. It should be noted that perturbations of the excitation condition from the target one $A_n$ and/or disorder in the system will introduce some deviations of $\mathcal{D}_s$ from the predicted value (8). As an example, the right panels in Fig.3 show the statistical distribution of $\mathcal{D}_s$ obtained when the array is excited with the perturbed amplitudes $a_n(0)=A_n ^{(S)}(1+f_n)$, where $f_n$ is taken from a Gaussian distribution with zero mean and $\sigma=0.1$ standard deviation.\\ 
\begin{figure}[htb]
\centerline{\includegraphics[width=8.3cm]{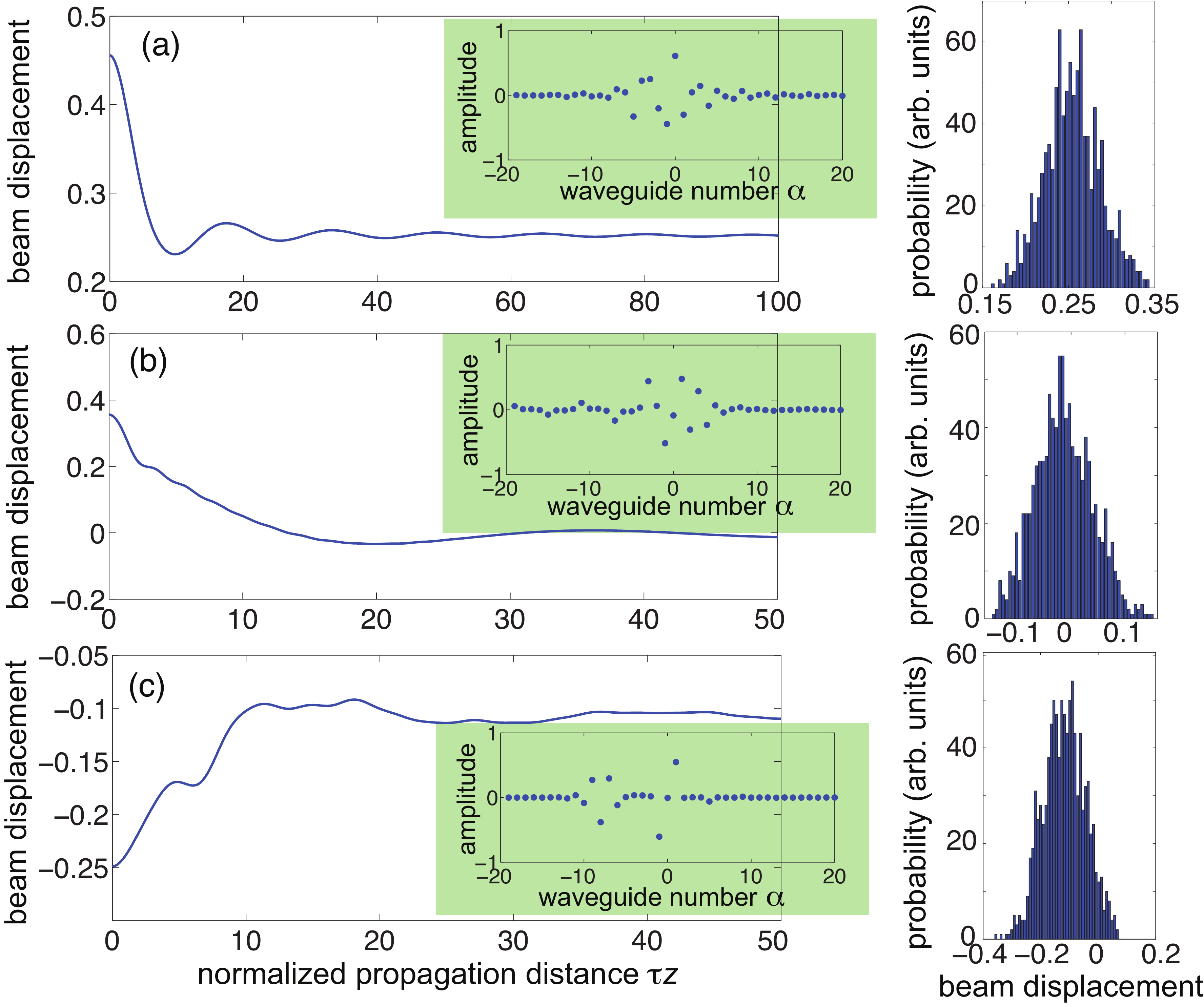}} \caption{ \small
(Color online) (a) Numerically-computed integral beam displacement $\mathcal{D}(z)$ versus normalized propagation distance $\tau z$ for the superlattice with inversion symmetry of Fig.2(a) and for $\tau=1$. Initial beam excitation, shown in the inset, corresponds to $S=2$ band excitation, i.e. $a_n(0)=A_n ^{(S)}$. The asymptotic value ${\mathcal D}_{as}$ approaches $\mathcal{N}_2 /(2 \pi S)=0.25$. The right panel shows the statistical distribution of $\mathcal{D}_{as}$ obtained for 1000 numerical runs corresponding to perturbed initial excitation amplitudes $a_n(0)=A_n^{(S)}(1+ f_n)$, where $f_n$ is a random number with Gaussian distribution of zero mean and $\sigma=0.1$ standard deviation. (b). Same as (a), but for $\tau=0.5$ and $S=3$ (three band excitation). In this case the  ${\mathcal D}_{as}=\mathcal{N}_3 /(2 \pi S)=0$.
(c) Same as (a), but for the superlattice of Fig.2(b) without inversion symmetry, $\tau=0.5$ and $S=3$ (excitation of the lowest three bands). In this case  ${\mathcal D}_{as}=\mathcal{N}_3 /(2 \pi S) \simeq -0.106$.}
\end{figure} 
{\it Bulk probing of zero-energy topological edge mode.} The Wannier function excitation protocol discussed above provides a rather general and powerful tool for topological spectroscopy of 1D superlattices. However, it requires the knowledge of Wannier functions, its practical implementation requires special beam shaping methods, and it is sensitive to perturbation of the excitation condition. It is thus worth exploring whether simpler excitation configurations can provide some useful information about the topological properties of the superlattice, without the knowledge of Wannier functions. The most interesting case is the one of a chiral superlattice ($M$ even), that can sustain a zero-energy mode protected by the particle-hole symmetry \cite{r27}. As shown in Fig.2(b), in a superlattice with broken inversion symmetry the topological zero-energy mode can coexist with other non-topological edge states in the other bang gaps. Clearly, edge excitation of the array -- like e.g. in the recent experiment \cite{r8} --can not discriminate in a simple way the existence of the topological zero-energy mode from other non-topological edge states, since in both cases some light remains trapped at the edge and oscillatory behavior is observed, as shown in Fig.4. For $M=4$, we found the following remarkable result. Let us consider single-cell excitation in the bulk corresponding to the initial condition $a_n=(1/ \sqrt{2}) (1, 0, \pm i, 0)^T \delta_{n,0}$. This means that only two waveguides of the array, separated apart by two sites, are equally excited at the input plane with a $\pm \pi/2$ phase shift; a possible excitation scheme is depicted in Fig.4(a). The spectral amplitudes $C_l$ corresponding to such an excitation condition are obtained from Eq.(5), and the corresponding asymptotic value of mean beam displacement $\mathcal{D}_{as}$ can be then calculated using Eq.(7). The final result is very simple: $\mathcal{D}_{as}=0$ for $\tau<\tau_0$ and $\mathcal{D}_{as}=-1/4$ for $\tau> \tau_0$, where $\tau_0 \equiv t_1 t_3 / t_2$ is the gap closing point above which the topological zero-energy mode does appear [Fig.2(b)]. This means that the bulk parameter $\mathcal{D}_{as}$ for the excitation $a_n(0)=(1/\sqrt{2})(1,0,\pm i,0)^T \delta_{n,0}$  provides a topological number that signals the existence (or not) of the topological zero-energy mode; see Figs.4(b) and (c). This topological number is analogous to the global winding number of the SSH$_4$ model recently introduced in \cite{r27bis}, however in our setup it is not required to trace the chiral (sublattice) average displacements over a basis of states of the unit cell, taken as different initial excitation conditions.
 \begin{figure}[htb]
\centerline{\includegraphics[width=8.3cm]{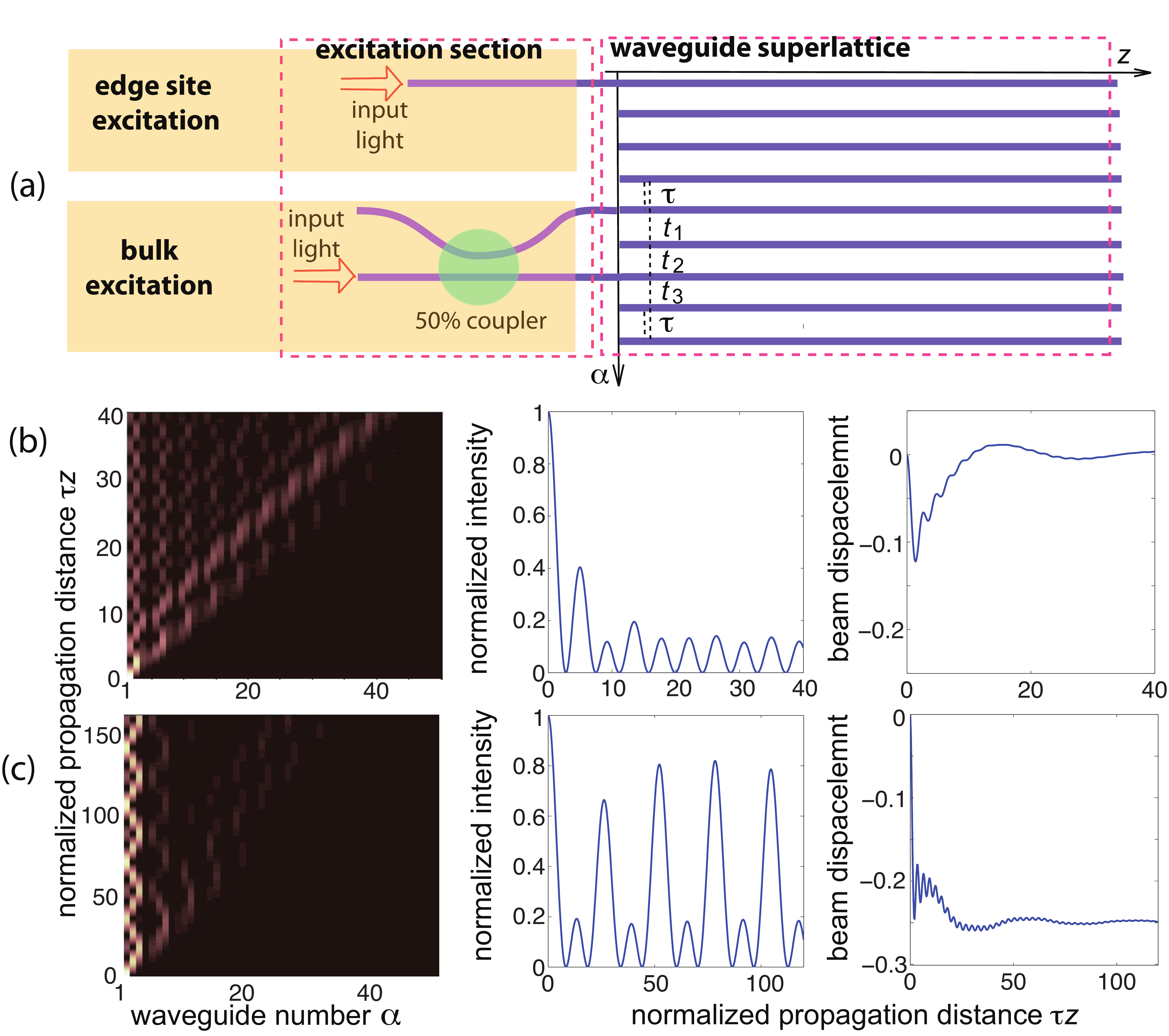}} \caption{ \small
(Color online) (a) Schematic of different types of excitation of the waveguide array (edge and bulk excitation). (b,c) Numerically-computed beam intensity evolution along normalized propagation distance for left-edge excitation (left panels), and corresponding evolution of the light intensity in the left edge waveguide (central panels) for the quadrimeric superlattice of Fig.2(b) with broken inversion symmetry. In (b) there are two non-topological left edge states that give rise to beating in the light intensity evolution, while in (c) two non-topological edge states coexist with the topological zero-energy edge state. The right panels show the behavior of the mean beam displacement $\mathcal{D}(z)$ for bulk excitation. Here two waveguides of the lattice are excited by waves of the same intensity and $\pm \pi /2$ shift using a balanced integrated optical directional coupler.  The four-port directional coupler is excited in one of the two input ports; for a $50 \%$ beam splitting ratio, at the two output ports light waves have the same intensity and a $ -\pi/2$ phase shift.  Parameter values are $t_1=0.3$, $t_2=0.2$, $t_3=0.5$, and $\tau=0.5$ in (b), $\tau=1.5$ in (c).}
\end{figure}

{\it Conclusions.} Identifying ways to relate global topological invariants with measurable quantities is a quest of major relevance in topological physics. Several methods have been suggested and demonstrated using synthetic topological matter. Here a general method of topological spectroscopy of 1D photonic superlattices has been suggested, which is based on bulk waveguide array excitation by a superposition of Wannier functions. In case of a quadrimeric superlattice, we also suggested a simple method to detect topological zero-energy edge states when they coexist with non-topological ones. Our results are feasible for a description of diverse 1D topological systems beyond optics, such as  electronic, polaritonic, mechanical, and acoustic superlattices. Future research directions include extension of  the Wannier function method to other systems, such as to topological non-Hermitian lattices, and improvement of the robustness against lattice disorder and perturbations.

%\begin{figure}[htbp]
 % \includegraphics[width=82mm]{Fig3.pdf}\\
  % \caption{(color online) Numerically-computed evolution of the occupation probabilities $P_1$ and $P_{2N-1}$ of left and edges sites in a SSH chain comprising $(2N-1)=31$ sites. Parameter values and the evolution of hopping amplitudes $t_1, t_1^{\prime}$ are as in Fig.2. The inset shows the evolution of the occupation probabilities of edge states $|L \rangle$ and $|R \rangle$ as predicted by the three-level CTAP model. The interaction time is $T=800$.}
%\end{figure}

%\clearpage
%\bibliography{H:/Physik/bibliography}
%\begin{thebibliography}{31}

%\end{thebibliography}
\end{document}